\documentclass[a4paper, 11pt]{article}

\usepackage{fullpage}
\usepackage[english]{babel}
\usepackage[latin1]{inputenc}
\usepackage{mathrsfs}
\usepackage{amsfonts}
\usepackage{amssymb}
\usepackage{amsmath}
\usepackage{amsthm}
\usepackage{graphicx}
\usepackage[usenames,dvipsnames]{color}
\usepackage[colorlinks=true,citecolor=blue,linkcolor=magenta]{hyperref}
\usepackage{lmodern}
\usepackage{microtype}
\usepackage{braket}
\usepackage{tabularx,colortbl}
\usepackage{pifont}
\usepackage{mathtools}
\usepackage{color}
\usepackage{dsfont}
\usepackage{algpascal}
\usepackage{algorithm} 
\usepackage{algpseudocode}
\usepackage{chngcntr}
\usepackage{complexity}
\usepackage{colortbl}
\usepackage{xfrac}
\usepackage[affil-it]{authblk}
\usepackage[table, x11names]{xcolor}
\usepackage{booktabs}
\usepackage{multirow}
\usepackage{tabulary}
\usepackage{array,makecell,multirow,ragged2e}
\usepackage[font=small,labelfont=bf]{caption}
\usepackage{csquotes}

\renewcommand{\thefootnote}{\fnsymbol{footnote}}

\renewcommand{\arraystretch}{1.5}

\newcommand{\comment}[1]{} 
 
\newcommand{\idty}[1]{\mathbb{1}} 
\newcommand{\ovsqrt}[1]{\frac{1}{\sqrt{2}}}
 
\newcommand{\tr}[1]{\mathrm{Tr}}
\newcommand{\Ord}[1]{\mathcal{O}(#1)}

\newcommand{\tOrd}[1]{\tilde{\mathcal{O}}(#1)}

\usepackage[backend=bibtex, natbib=true, bibstyle=alphabetic,citestyle=alphabetic, doi=true, sorting=nyt, firstinits]{biblatex}
\title{Quantum machine learning: a classical perspective}

\begin{document}

\author[1]{Carlo Ciliberto}
\author[1]{Mark Herbster}
\author[2,3]{Alessandro Davide Ialongo}
\author[1,4]{Massimiliano Pontil}
\author[1,5]{Andrea Rocchetto \thanks{Corresponding author: andrea.rocchetto@spc.ox.ac.uk}}
\author[1,6]{Simone Severini}
\author[1,5,7]{Leonard Wossnig}

\affil[1]{Department of Computer Science, University College London}
\affil[2]{Department of Engineering, University of Cambridge}
\affil[3]{Max Planck Institute for Intelligent Systems, T\"{u}bingen}
\affil[4]{Computational Statistics and Machine Learning - Istituto Italiano di Tecnologia}
\affil[5]{Department of Materials, University of Oxford}
\affil[6]{Institute of Natural Sciences, Shanghai Jiao Tong University}
\affil[7]{Theoretische Physik, ETH Z\"{u}rich}

\date{}
\maketitle
\renewcommand{\thefootnote}{\arabic{footnote}}

\begin{abstract}

Recently, increased computational power and data availability, as well as algorithmic advances, have led machine learning techniques to impressive results in regression, classification, data-generation and reinforcement learning tasks. Despite these successes, the proximity to the physical limits of chip fabrication alongside the increasing size of datasets are motivating a growing number of researchers to explore the possibility of harnessing the power of quantum computation to speed-up classical machine learning algorithms. Here we review the literature in quantum machine learning and discuss perspectives for a mixed readership of classical machine learning and quantum computation experts. Particular emphasis will be placed on clarifying the limitations of quantum algorithms, how they compare with their best classical counterparts and why quantum resources are expected to provide advantages for learning problems. Learning in the presence of noise and certain computationally hard problems in machine learning are identified as promising directions for the field. Practical questions, like how to upload classical data into quantum form, will also be addressed.\\
\end{abstract}

\hypersetup{linkcolor=blue}

\tableofcontents


\section{Introduction}

In the last twenty years, thanks to increased computational power and the availability of vast amounts of data, \textit{machine learning} (ML) algorithms have achieved remarkable successes in tasks ranging from computer vision~\cite{krizhevsky2012imagenet} to playing complex games such as Go~\cite{silver2016mastering}. However, this revolution is beginning to face increasing challenges. With the size of datasets constantly growing and Moore's law coming to an end~\cite{markov2014limits}, we might soon reach a point where the current computational tools will no longer be sufficient. Although tailored hardware architectures, like graphics processing units (GPUs) and tensor processing units (TPUs), can significantly improve performance, they might not offer a structural solution to the problem.\\

Quantum computation is a computational paradigm based on the laws of quantum mechanics. By carefully exploiting quantum effects like interference or (potentially) entanglement, quantum computers can efficiently solve selected problems~\cite{shor1997polynomial, van2006quantum, hallgren2007polynomial} that are believed to be hard for classical machines. This review covers the intersection of machine learning and quantum computation, also known as \textit{quantum machine learning} (QML).
The term quantum machine learning has been used to denote different lines of research such as using machine learning techniques to analyse the output of quantum processes or the design of classical machine learning algorithms inspired by quantum structures. For the purpose of this review we refer to QML solely to describe learning models that make use of quantum resources.\\

The goal of this review is to summarise the major advances in QML for a mixed audience of experts in machine learning and quantum computation and serve as a bridge between the two communities. Most problems will be analysed under the lens of computational complexity, possibly, a unifying language for both communities. We do not aim for completeness but rather discuss only the most relevant results in quantum algorithms for learning. For the interested reader there is now a number of resources covering quantum machine learning in the broader sense of the term~\cite{adcock2015advances,biamonte2016quantum}. For an introduction to quantum algorithms we refer to the reviews of Montanaro~\cite{montanaro2016quantum} and Bacon~\cite{bacon2010recent}, while for machine learning to the books by Bishop~\cite{bishop2006pattern} and Murphy~\cite{murphy2012machine}.\\ 

Why should a machine learning expert be interested in quantum computation? And why are we expecting quantum computers to be useful in machine learning? We can offer two reasons. First, with an ever growing amount of data, current machine learning systems are rapidly approaching the limits of classical computational models. In this sense, quantum algorithms offer faster solutions to process information for selected classes of problems. Second, results in quantum learning theory point, under certain assumptions, to a provable separation between classical and quantum learnability. This implies that hard classical problems might benefit significantly from the adoption of quantum-based computational paradigms. But optimism should come with a dose of scepticism. The known quantum algorithms for machine learning problems suffer from a number of caveats that limit their practical applicability and, to date, it is not yet possible to conclude that quantum methods will have a significant impact in machine learning. We will cover these caveats in detail and discuss how classical algorithms perform in light of the same assumptions. \\ 

Quantum computation is a rapidly evolving field but the overarching question remains: when will we have a quantum computer? Although it is not within the scope of this review to present a time-line for quantum computation it is worth noting that in the last years the worldwide effort to build a quantum computer has gained considerable momentum thanks to the support of governments, corporations and academic institutions. It is now the general consensus that general purpose
quantum computation is within a $15$ years time-line~\cite{de2016quantum}.\\

The review is structured as follow. We start with Section~\ref{sec:essential} by providing some essential concepts in quantum computation for the reader with no prior knowledge of the field. In Section~\ref{sec:setting} we introduce the standard models of learning, their major challenges and how they can be addressed using quantum computation. Section~\ref{sec:better} surveys results in quantum learning theory that justify why we expect quantum computation to help in selected learning problems. We proceed in Section~\ref{sec:data} by discussing how to access data with a quantum computer and how these models compare with parallel architectures. We continue by presenting different computational and mathematical techniques, that find widespread application in machine learning, and can be accelerated with a quantum computer. More specifically, we survey quantum subroutines to speedup linear algebra (Section~\ref{sec:linear}), sampling (Section~\ref{sec:sampling}), and optimisation problems (Section~\ref{sec:opt}). For each section, we discuss the asymptotic scaling of the classical and quantum subroutine and present some learning applications. The following section is dedicated to quantum neural networks (Section~\ref{sec:QNN}). Even if neural networks are not a mathematical technique on their own, they are surveyed in a dedicated section due to their prominence in modern machine learning. The last two sections cover two promising applications of quantum computation in machine learning. In Section~\ref{sec:noise} we consider the case of learning under noise while in Section~\ref{sec:hard} we discuss computationally hard problems in machine learning. We conclude with an outlook section.

\section{Essential quantum computation}
\label{sec:essential}

Quantum computing focuses on studying the problem of storing, processing and transferring information encoded in quantum mechanical systems. This \emph{mode} of information is consequently called \emph{quantum information}. The book by Nielsen and Chuang~\cite{nielsen2010quantum} is a standard introduction to the field. Loosely speaking, quantum computational models propose a probabilistic version of (time) reversible computation, \textit{i.e.} computation in which the output is in one-to-one correspondence with the input. According to quantum theory, physical states are mathematically represented by density matrices, which are trace-one, positive-semidefinite matrices that generalise the concept of probability distributions. The logical states used by a quantum computational model are then identified with the physical states of the quantum system that implements it.  A computation is executed reversibly by applying a sequence of unitary matrices to an initialised state. A probabilistic output is obtained according to the distribution encoded by the final density matrix.\\

In this framework, the fundamental unit of quantum information is the state of any quantum system with two degrees of freedom distinguishable by an observer, which then coincide with the usual logical values $0$ and $1$. This is called \emph{qubit} and, for our purposes, it is a vector $\psi = \alpha_0 e_0 + \alpha_1 e_1$, where $\alpha_0, \alpha_1 \in \mathbb{C}$, $|\alpha_0|^2 + |\alpha_1|^2 = 1$ and, $e_i$ denotes the $i$-th standard basis vector. The values of a distribution on $0$ and $1$ are given by $\{|\alpha_1|^2,|\alpha_2|^2\}$. It is a basic fact that the information content of a qubit is equivalent to a single bit. Registers of multiple qubits are assembled with the use of a tensor product. Unitary matrices acting on a small number of qubits can then be interpreted as a generalisation of logic gates. The induced dynamics is responsible for interference, a key ingredient of quantum computation. By exploiting interference effects, quantum computers are able to simultaneously evaluate a function on every point of its domain. Although the result of this computation is not immediately accessible to a classical observer, the possibility of using quantum dynamics to increase the probability of determining a given property of the function is promised to allow a quantum computer to solve some computational problems exponentially faster than classical devices. Although the true roots, and extents, of quantum speedups are still unclear, it is believed that structure, certain symmetries and non-classical correlations play an important role in the observed advantages~\cite{jozsa2003role, aaronson2009need}. \\

In the context of the analysis of classical data, we can exploit the encoding of quantum information to efficiently represent classical probability distributions with exponentially many points. For instance, when $v = (v_1, \dots, v_{2^n})$ is a probability vector of size $2^n$, we can write an $n$-qubit state (register), $ \psi = \sum_{i=1} ^{2^n} \sqrt{v_i} e_i $.\\

Finally, when we use the term qubits we always refer to idealised, error free, objects. In practice quantum states are extremely fragile and require extensive error correction to be shielded from the effects of noise. The theory of error correction guarantees that if the physical errors are below a certain threshold it is possible to correct the system efficiently. The theory of error correction is reviewed by Preskill~\cite{preskill1998fault}. We further discuss the types of error affecting quantum systems in Section~\ref{sec:noise}.

\subsection*{Comparing the performance of classical and quantum algorithms}

Computational complexity studies the resources needed to perform a given computational task. One of the major goals of this theory is to classify problems according to their time complexity, which roughly corresponds to the number of elementary steps necessary to solve the problem as a function of the size of the input. The books by Papadimitrou~\cite{papadimitriou2003computational} and Arora and Barak~\cite{arora2009computational} provide extensive introductions. \\

We define \textit{quantum speedup} as the advantage in runtime obtained by a quantum algorithm over the classical methods for the same task. We quantify the runtime with the asymptotic scaling of the number of elementary operations used by the algorithm with respect to the size of the input. To compare the performance of algorithms, we use the computer science notation $\mathcal{O}(f(n))$ indicating that the asymptotic scaling of the algorithm is upper-bounded by a function $f(n)$ of the number $n$ of parameters characterizing the problem, \textit{i.e.} the size of the input. The notation $\tilde{\mathcal{O}}(f(n))$ ignores logarithmic factors. \\

In computational complexity theory it is customary to analyse algorithms with respect to the number of queries to some efficiently implementable oracles, which can be either classical or quantum. This approach to analysing algorithms is called the \textit{query model}. In the query model an algorithm is said to be \textit{efficient} if it queries the oracle a polynomial number of times. Throughout this review many speedups are obtained in the query model. A standard oracle for QML assumes that the classical datasets can be efficiently produced in quantum superposition. The QRAM discussed in Section~\ref{sec:data} is a possible implementation of this oracle.

\section{Setting the problem: perspectives in machine learning} 
\label{sec:setting}

The term machine learning refers to a variety of statistical methods to analyse data. The prototypical goal is to infer, from a finite number of observations (the training data), the future behaviour of an unknown and possibly non-deterministic process (such as the dynamics of the stock market or the activations patterns in the human brain). In the last four decades, the field of machine learning has grown to such an extent that even providing a brief overview of the most prominent ideas and frameworks would require a review on its own. To this end, for the purpose of this review, we mainly focus on one of the most well-established and mature areas of research, namely supervised learning from the perspective of learning theory. To the reader interested in a more satisfying overview of machine learning we recommend, for instance, \cite{murphy2012machine}.\\

Learning theory aims to place the problem of learning from data on solid mathematical foundations. Typical questions that one asks in this setting are: how many examples are required to learn a given function? How much computational resources are required to perform a learning task? Depending on a number of assumptions about the data access model and on the goal of learning, it is possible to define different learning models. Two prominent ones are the {\em Probably Approximately Correct (PAC)} framework developed by Valiant~\cite{valiant1984theory} and the {\em Statistical Learning Theory} by Vapnik~\cite{vapnik1998}. Here, a learner seeks to approximate an unknown function based on a training set of input-output pairs. Examples in the training set are assumed to be drawn from an unknown probability distribution and predictions are tested on points drawn from the same distribution. PAC and statistical learning theory model the efficiency of an estimator with two quantities: the sample complexity and the time complexity. The \textit{sample complexity} is the minimum number of examples required to learn a function up to some approximation parameters and it is directly related to the capacity of the hypotheses space and the regularity of the data distribution; the \textit{time complexity} corresponds to the runtime of the best learning algorithm. A learning algorithm is said to be efficient if its runtime is polynomial in the dimension of the elements of the domain of the function and inverse polynomial in the error parameters.\\    

In these settings the goal is to find a model that fits well a set of training examples but that, more importantly, guarantees good prediction performance on new observations. This latter property, also known as {\em generalisation capability} of the learned model, is a key aspect separating machine learning from the standard optimisation literature. Indeed, while data fitting is often approached as an optimisation problem in practice, the focus of machine learning is to design statistical estimators able to ``fit'' well future examples. This question is typically addressed with so-called {\em regularisation} techniques, which essentially limit the expressive power of the learned estimator in order to avoid overfitting the training dataset.\\

A variety of regularisation strategies have been proposed in the literature, each adopting a different perspective on the problem (see~\cite{vapnik1998,bishop2006pattern,bauer2007regularization} for an introduction on the main ideas). Among the most well-established approaches it is worth mentioning those that directly impose constraints on the hypotheses class of candidate predictors (either in the form of hard constraints or as a penalty term on the model parameters, such as in Tikhonov regularisation) or those that introduce the regularisation effect by ``injecting'' noise in the problem (see Section~\ref{sec:noise}). These ideas have led to popular machine learning approaches currently used in practice, such as Regularised Least Squares~\cite{cucker2002}, Gaussian Process (GP) Regression and Classification \cite{rasmussen2006}, Logistic Regression~\cite{bishop2006pattern}, and Support Vector Machines (SVM)~\cite{vapnik1998} to name a few.\\

From a computational perspective, regularisation-based methods leverage on optimisation techniques to find a solution for the learning problem and typically consist of a sequence of standard linear algebra operations such as matrix multiplication and inversion. In particular, most classical algorithms, such as GP or SVM, require a number of operations comparable to that of inverting a square matrix that has size equal to the number $N$ of examples in the training set. This leads, in general, to a time complexity of $\mathcal O(N^3)$ which can be improved depending on the sparsity and the conditioning of the specific optimisation problem, see Section~\ref{sec:linear}. However, as the size of modern datasets increases, the above methods are approaching the limits of their practical applicability. \\

Recently, alternative regularisation strategies have been proposed to reduce the computational costs of learning. Instead of considering the optimisation problem as a separate process from the statistical one, these methods hinge on the intuition that reducing the computational burden of the learning algorithm can be interpreted as a form of regularisation on its own. For instance, {\em early stopping} approaches perform only a limited number of steps of an iterative optimisation algorithm (such as gradient descent) to avoid overfitting the training set. This strategy clearly entails less operations (less number of steps) but can be shown theoretically to lead to same generalisation performance of approaches such as Tikhonov regularisation~\cite{bauer2007regularization}. A different approach, also known as {\em divide and conquer}, is based on the idea of distributing portions of the training data onto separate machines, each solving a smaller learning problem, and then combining individual predictors into a joint one. This strategy benefits computationally from both the parallelisation and reduced dimension of distributed datasets and it has been shown to achieve the same statistical guarantees of classical methods under suitable partitions of the training data \cite{zhang2013divide}. A third approach that has recently received significant attention from the machine learning community is based on the idea of constraining the learning problem to a small set of candidate predictors, obtained by randomly sampling directions in a larger, universal hypotheses space (namely a space dense in the space of continuous function). Depending on how such sampling is performed, different methods have been proposed, the most well-known being random features~\cite{rahimi2007random} and Nystrom approaches~\cite{smola2000sparse,williams2000using}. The smaller dimensionality of the hypotheses space automatically provides an improvement in computational complexity. It has been recently shown that it is possible to obtain equivalent generalisation performance to classical methods also in these settings \cite{rudi2015less}.\\

For all these methods, training times can be typically reduced from the $\mathcal O(N^3)$ of standard approaches to $\widetilde{ \mathcal{O}} (N^2)$ while keeping the statistical performance of the learned estimator essentially unaltered. \\

Because the size of modern datasets is constantly increasing, time complexities in the order of $\widetilde{ \mathcal{O}} (N^2)$ might still be too demanding for practical applications. In this regard, quantum computation could offer the potential to further improve the efficiency of such methods allowing them to scale up significantly. Indeed, through a number of quantum algorithms for linear algebra, sampling and optimisation techniques, we could in principle obtain up to exponential speedups over classical methods. However, as it will be discussed in Section~\ref{sec:linear}, current QML methods require fast memory access and particular data structures that might limit their applicability in real settings. Nevertheless, as we will discuss in the following section, a number of results in quantum learning theory point, under specific assumptions, to a clear separation between classical and quantum learning paradigms in some specific settings.

\section{``Can we do better?'': insights from quantum learning theory}
\label{sec:better}

Learning theorists have been interested to study how quantum resources can affect the efficiency of a learner since the $90$'s. Although different learning models have been translated into the quantum realm, here we focus on the quantum version of the PAC model. The reason of this choice is that in this model we have results for both the sample and the time complexity. For an extensive overview of the known results in quantum learning theory we refer the reader to the review by Arunachalam and de Wolf~\cite{arunachalam2017survey}.\\ 

The quantum PAC model has been introduced in~\cite{bshouty1998learning}. Here it is assumed that the learner has access to a quantum computer and to an oracle that returns the training set in quantum superposition. 
In terms of sample complexity, it has been shown in a series of papers, which constantly improved the bounds until reaching provable optimality~\cite{servedio2004equivalences,atici2005improved,zhang2010improved,arunachalam2016optimal}, that the quantum PAC model under an unknown distribution and standard PAC are equivalent up to constant factors. This implies that, in general, quantum mechanics does not help to reduce the amount of data required to perform a learning task. However, if one considers a different learning model, like the exact learning framework developed by Angluin~\cite{angluin1988queries}, it is possible to prove that quantum learners can be polynomially more efficient than classical in terms of number of queries to the data oracle~\cite{bshouty1994oracles,servedio2004equivalences}. \\

Although quantum and classical examples are equivalent up to constant factors when learning under general distributions, the quantum PAC model can offer advantages over its classical counterpart in terms of time complexity. One of the central problems studied in the classical literature is the learnability of disjunctive normal forms (DNFs). To date, the time complexity of the best algorithm for learning DNFs under an unknown distribution is exponential~\cite{klivans2001learning}. A number of assumptions can be made to relax the hardness of the problem. For instance, if the learner is provided with examples drawn from the uniform distribution then the runtime of the best learner becomes quasipolynomial~\cite{verbeurgt1990learning}. For the sake of completeness we note that the methods presented in Section~\ref{sec:setting}, like SVMs, have been shown to not being able to learn efficiently DNF formulas~\cite{ben2002limitations, khardon2005maximum}. The learnability of DNF formulas has also been studied in the quantum PAC model~\cite{bshouty1998learning}. Here DNFs have been shown to be efficiently learnable under the uniform distribution. This quantum speedup is obtained through an efficient algorithm~\cite{bernstein1997quantum} that allows to sample exponentially faster from the probability distribution described by the coefficients of a boolean Fourier transform. Interestingly DNF formulas can be shown to be efficiently learnable under noise. We will return to this point in Section~\ref{sec:noise}.\\

Another case where it is believed that learning can be performed efficiently only when the learner has access to quantum resources is based on a class of functions developed by Kearns and Valiant~\cite{kearns1994cryptographic}. This class is provably hard to learn under the assumption that factoring Blum integers is also hard (an assumption widely believed to be true; for a brief introduction to the concept of hardness in computational complexity see Section~\ref{sec:hard}). Servedio and Gotler~\cite{servedio2004equivalences}  noted that thanks to Shor's  quantum factoring algorithm~\cite{shor1997polynomial} this class of functions can be learned efficiently in the quantum PAC model.\\

The results coming from the quantum learning theory literature show that by carefully exploiting quantum mechanical effects, depending on the type of learning model considered, it is possible to have a better generalisation error (\textit{i.e.} we can learn with less examples) or we can learn functions that would otherwise be hard for classical learners.

\section{Data access, communication and parallelism}
\label{sec:data}

One of the roots of the speedups theorised in quantum computation is the ability to process information in quantum superposition~\cite{nielsen2010quantum,aaronson2013quantum}. Because machine learning is ultimately about analysing vast amounts of data it is important to address the question of how data is turned into quantum superposition. We distinguish between two types of algorithms: those that operate on quantum data (\textit{i.e.} data that is output of a quantum process, for example, a quantum chemistry problem) and those that seek to process data stored in a classical memory. The first case is ideal for QML. The data is ready to be analysed and we do not have to spend computational resources to convert the data into quantum form. The second case is more elaborate as it requires a procedure that encodes the classical information into a quantum state. As we will see the computational cost of this operation is particularly relevant to determine whether we can obtain quantum speedups in machine learning for classical data.\\

Let us assume that one wants to process $N$ $d$-dimensional classical vectors with a quantum algorithm. The \textit{quantum random access memory} (QRAM)~\cite{giovannetti2008qram1,giovannetti2008qram2} is  a quantum device that can encode in superposition $N$ $d$-dimensional vectors into $\mathrm{log}(Nd)$ qubits in $\mathcal{O}(\mathrm{log}(Nd))$ time by making use of the so called ``bucket-brigade'' architecture. 
The idea is to use a tree-structure where the $Nd$ leaves contain the entries of the N vectors in $\mathbb{R}^d$. The QRAM, with a runtime of  $\Ord{\mathrm{log}(Nd)}$, can return a classical vector in quantum superposition efficiently. However, the number of physical resources it requires scales as $\Ord{Nd}$. As we will see this exponential scaling (with respect to the number of qubits) has been used to question whether the QRAM can be built in an experimental setting or whether it can provide a genuine computational advantage~\cite{aaronson2015read,adcock2015advances}. Fundamentally the issue can be related to whether the exponential number of components needs to be continuously ``active''. The proponents of the QRAM~\cite{giovannetti2008qram1,giovannetti2008qram2} claim that only $\Ord{\mathrm{log}(Nd)}$ components need to be active while the others can be considered as ``non-active'' and error free. Whether this assumption holds in an experimental setting is unclear~\cite{arunachalam2015robustness}. We now proceed to discuss its implications.\\

The first issue that appears with QRAM is whether all the components require to be error corrected (we briefly discuss errors in quantum computation in Section~\ref{sec:noise}). Indeed, if the exponential physical resources required full error correction then it would be impractical to build the device in an experimental setting. Arunachalam \textit{et al.}~\cite{arunachalam2015robustness} addressed this question and showed that, provided a certain error model, algorithms that require to query the memory a polynomial number of times (like the quantum linear system algorithm presented in Section~\ref{sec:linear}) might not require fault-tolerant components. However, for superpolynomial query algorithms, like Grover's search~\cite{grover1996fast} (a subroutine required, for example, in some of the quantum methods for training restricted Boltzmann machines discussed in Section~\ref{sec:QNN}) the QRAM requires error-corrected components.\\

A second problem related to the exponential number of resources in an active memory has been raised by Aaronson~\cite{aaronson2015read} and by Steiger and Troyer~\cite{steiger2016perimeter}. The authors argue that the only fair comparison of a system which requires an exponential number of resources is with a parallel architecture with a similar amount of processors. In this case many linear algebra routines, including solving linear systems and singular value decomposition, can be solved in logarithmic time~\cite{csanky1976fast}).\\

A third caveats of the QRAM is the requirement of having data distributed in a relatively uniform manner over the quantum register. As pointed out in~\cite{prakash2014quantum, aaronson2015read}, if that was not the case the QRAM would violate the search lower bounds proved in~\cite{bennett1997strengths}. In the case of non-uniformly distributed data, the QRAM is no longer efficient and take $\Ord{\sqrt{N}}$ to turn the classical dataset into quantum superposition.\\

As a last comment on the QRAM, the possibility of loading the data in logarithmic time, when the size of the data is considerable, can be controversial due to speed of communication arguments. In fact, as suggested in~\cite{aaronson2015read}, latency can play a role in big memory structures. In particular, a lower bound on the distance which the information has to travel implies a lower bound on latency, due to considerations on the limits set by the speed of light. In a three dimensional space these are given by $\Ord{\sqrt[3]{Nd}}$. In practice these considerations will only dominate if the amount of memory is extremely large but, because in quantum machine learning we aim at datasets that surpass the current capability of classical computers, this bound is a potential caveat.\\


In conclusion, the QRAM allows to upload data efficiently but might be hard to implement experimentally or might not allow a genuine quantum advantage if we take into account all the required resources. Noticeably the fast data access guaranteed by the QRAM is only required for QLM algorithm that run in sublinear time. Although many known QML algorithms run in sublinear time, quantum learning theory suggest that for some classically hard problems quantum resources might give exponential advantages. In this case, a memory structure that can prepare a quantum superposition in polynomial time (\textit{i.e.} in $\mathcal{O}(Nd)$) can still be sufficient to maintain a quantum speedup compared to the classical runtime. We will discuss hard classical learning problem in Section~\ref{sec:hard}.

Finally, we note that, although the QRAM, due to its generality, is the most widely used memory structure in QML algorithms other protocols to encode classical data in superposition exist. For example, a technique developed by Grover and Rudolph allows one to generate a quantum superposition that encodes an approximate version of a classical probability distribution provided its density is efficiently integrable~\cite{grover2002creating}.

\section{Fast linear algebra with quantum mechanics}
\label{sec:linear}

A significant number of methods in the quantum machine learning literature is based on fast quantum algorithms for linear algebra. In this section we present the two main quantum subroutines for linear algebra: a quantum algorithm for matrix inversion and a quantum algorithm for singular value decomposition. We summarise the major applications of these techniques to machine learning problems and how they compare with classical an parallel implementations in Table~\ref{tab:QMLlinalgebra}.

\subsection*{Fast matrix inversion: the quantum linear system algorithm}

Solving linear systems of equation is an ubiquitous problem in machine learning. As discussed in Section~\ref{sec:setting}, many learning problems, like Gaussian Processes or SVMs, require the inversion of a matrix. For a system of linear equations $A x = b$ with $A \in \mathbb{R}^{N \times N}$ and $x,b \in \mathbb{R}^N$, the best classical algorithm has a runtime of $\Ord{N^{2.373}}$~\cite{coppersmith1990matrix}. However, due to a large pre-factor, the algorithm is not used in practice. Standard methods, for example, based on QR-factorisation take $\Ord{N^3}$ steps \cite{golub2012matrix}.\\

The \textit{quantum linear system algorithm} (QLSA)~\cite{harrow2009quantum}, also known as HHL after the three authors Harrow, Hassidim, and Lloyd, promises to solve the problem in $\tOrd{\log (N) \kappa^{2} s^{2} /\epsilon}$, where $\kappa$ is the condition number (defined to be the ratio of the largest to the smallest eigenvalue), $s$ is the sparsity or the maximum number of non-zero entries in a row and column of $A$ and $\epsilon$ is the precision to which the solution is approximated. The precision is defined as the distance of the solution vector $\overline{a}$ to the true result $a$, which is given by $|| \overline{ a} -  a || = \sqrt{1-2 \text{Re} \langle \overline{ a},  a \rangle} \leq \epsilon$. Ambainis~\cite{ambainis2012variable} and Childs \textit{et al.}~\cite{childs2015quantum} improved the runtime dependency of the algorithm in $\kappa$ and $s$ to linear and the dependency in $\epsilon$ to poly-logarithmic.\\

Although the QLSA algorithm solves matrix inversion in logarithmic time a number of caveats might limit its applicability to practical problems~\cite{aaronson2015read}. First, the QLSA algorithm requires the matrix $A$ to be sparse. Second, the classical data must be loaded in quantum superposition in logarithmic time. Third, the output of the algorithm is not $x$ itself but a quantum state that encodes the entries of $x$ in superposition. Fourth, the condition number must scale at most sublinearly with $N$. An interesting problem that satisfies this requirements is discussed in~\cite{clader2013preconditioned}.\\ 

Recently, \cite{wossnig2017quantum} addressed the first caveat. By using a quantum walk based approach the authors derived an algorithm that scales as $\tOrd{\kappa^{2} ||A||_F \log N/\epsilon}$ and can also be applied to dense matrices (however, in this case, the speedup is only quadratic since $||A||_F=\sqrt{N}$). This result has been improved  in~\cite{kerenidis2017quantum}, currently the best known lower bound for matrices with this property.\\

The second caveat inherits the same issues of the QRAM discussed in Section~\ref{sec:data}: it is an open question whether we can practically load classical data in quantum superposition in logarithmic time. \\

The third caveat is a common pitfall of quantum algorithms. As pointed out by Childs~\cite{childs2009quantum} and Aaronson~\cite{aaronson2015read}, in order to  retrieve classical information from the quantum state, we need at least a number of measurements that is proportional to $N$. This would destroy every exponential speedup. One way forward is to use the QLSA algorithm only to compute certain features of the classical vector, which can be extracted efficiently using quantum mechanics, for example, the expected value $x^T A x$ of a matrix $A$. A number of other possible applications is discussed in~\cite{aaronson2015read}.\\

It is then natural to question how quantum algorithms compare to their classical analogues after all the caveats have been taken into account. For example, it is possible to show that calculating an expectation value of the form $x^T A x$ can be done in time linear in the sparsity of the matrix $A$, using classical sampling methods. Furthermore, conjugated gradient descent can obtain the full solution of the linear system (also when $A$ is not sparse) in linear time in the dimensionality and, in most cases, the sparsity of $A$~\cite{shewchuk1994introduction}. We present a general comparison of the asymptotic scalings of classical, quantum and parallel algorithms for linear algebra and their major applications in machine learning in Table~\ref{tab:QMLlinalgebra}.\\

\newcommand{\specialcell}[2][l]{\begin{tabular}[#1]{@{}l@{}}#2\end{tabular}}
\newcommand{\ra}[1]{\renewcommand{\arraystretch}{#1}}

\ra{1.5}
\begin{table}[t]
\centering
\small
\begin{tabulary}{\textwidth}{lll}
\rowcolor{black!20} \textbf{Problem} &  Solving linear system of equations & Singular value estimation  \\
\toprule
\toprule
\textbf{Scaling} &  
 \specialcell[l]{
C: $\tOrd{\sqrt{\kappa} \hat{s}N \log{(1/\epsilon)} }$~\cite{shewchuk1994introduction}\textsuperscript{\textbf{a}},\\
Q: $\tOrd{s^2 \kappa^2 \log{(N)} /\epsilon}$ ~\cite{harrow2009quantum}\textsuperscript{\textbf{b}}, \\
P: $\Ord{\log^2(N) \log{(1/\epsilon)}}$~\cite{csanky1976fast}\textsuperscript{\textbf{c}}
} & 
\specialcell[l]{
C: $\Ord{k^2 N \log{(1/\delta)}/\epsilon}$\cite{frieze2004fast}\textsuperscript{\textbf{d}},\\
Q: $\Ord{ \log{(N)} \epsilon^{-3}}$\cite{lloyd2014pca},\\
P: $\Ord{\log^2(N) \log{(1/\epsilon)}}$~\cite{csanky1976fast}\textsuperscript{\textbf{e}}} \\
\midrule
\textbf{Applications} & 
\specialcell{
Least-square-SVM~\cite{rebentrost2014quantum},\\GP Regression~\cite{zhao2015quantum},\\ Kernel Least Squares~\cite{schuld2016prediction}
} &
\specialcell{
Recommendation Systems~\cite{kerenidis2017quantum},\\Linear Regression~\cite{schuld2016prediction},\\Principal Component Analysis~\cite{lloyd2014pca}\textsuperscript{\textbf{f}}
}
\end{tabulary}
\caption{\textbf{Quantum linear algebra algorithms and their machine learning applications.} When carefully compared with classical versions that take into account the same caveats, quantum algorithms might loose their advantages.  C,Q,P indicate, respectively, the asymptotic computational complexity for classical, quantum, and parallel computation. We remind the reader that, to date, memory and bandwidth limits in the communication between processors make the implementation of certain parallel algorithms unrealistic. We remark that asymptotic scalings are only an indication of potential runtime differences and solely by benchmarking the algorithms on quantum hardware we will obtain clear insights on their performance. Given an $N\times N$ dimensional matrix $A$, we denote by $k$ the number of singular values that are computed by the algorithm, by $s$ the sparsity, and by $\kappa$ the condition number. For approximation algorithms $\epsilon$ is an approximation parameter. In other cases it denotes the numerical precision. Classical algorithms return the whole solution vector. Quantum algorithms return a quantum state; in order to extract the classical vector one needs $\mathcal{O}(N)$ copies on the state. \textbf{a)} is an approximate algorithm and can be applied to dense matrices. Here $\hat{s}N$ is the number of entries in the matrix $A$ and $\hat{s}$ alludes to the number of entries per row here; \textbf{b)} is exact but it does not output the solution vector and works only for sparse matrices (more details can be found in Section~\ref{sec:linear}); \textbf{c)} requires $\mathcal{O}(N^4)$ parallel units and it is numerically unstable due to high sensitivity to rounding errors. Stable algorithms such as Gaussian elimination with pivoting or parallel QR-decomposition require $\Ord{N}$ time using $\Ord{N^2}$ computational units~\cite{cosnard1986complexity}; \textbf{d)} is an approximate algorithm which returns a rank $k$-approximation with probability $1-\delta$, and has an additional error $\epsilon ||A||_F$. Exacts methods for an $N \times M$ matrix scale with $\min \{ MN^2,NM^2\}$; \textbf{e)} calculates SVD by computing the eigenvalue decomposition of the symmetric matrix $AA^T$; \textbf{f)} works on dense matrices that are low-rank approximable. Finally, we note that there exists efficient, classical, parallel algorithms for sparse systems, where $s=\Ord{\log N}$~\cite{li2009fast,li2012extension}. Probabilistic numerical linear algebra also allows to solve selected problems more efficiently and, under specific assumptions, even in linear time and with bounded error~\cite{halko2011finding}.
\label{tab:QMLlinalgebra}}
\end{table}  

Comparing algorithms based on their worst case running time may not be the right approach when considering their practical applicability, as it is commonly done in machine learning. Indeed, despite its worst case running time, an algorithm solving a given problem will often terminate much faster: average-case complexity can be much lower than worst case. Furthermore, \emph{smoothed analysis}~\cite{spielman2004smoothed, spielman2009smoothed} provides a framework for studying the time performance of an algorithm in the presence of realistic input noise distributions. This gives another way to quantify the complexity of algorithms. To date, no quantum algorithm has been analysed in the smoothed analysis framework.\\ 

Statistical considerations can also lead to interesting insights on the computational hardness of a learning problem. Kernel regularized least squares provide a good example. Under standard technical assumptions on the target function of the learning problem, computational regularization methods for kernel regularized least squares~\cite{bauer2007regularization,duchi2010,rudi2015less} (see Section~\ref{sec:setting}) achieve the optimal learning rate of $\epsilon = \Ord{N^{-1/2}}$ while requiring only $\tOrd{N^2}$ operations. With optimal learning rates we mean that any learning algorithm cannot achieve better prediction performance (uniformly) on the class of problems considered. Interestingly, such assumptions also allow us to derive estimates for the condition number of the kernel matrix to be of order $\kappa = \Ord{N^{1/2}}$ \cite{caponnetto2007}. The corresponding quantum scaling for the inversion of the kernel matrix is $\tOrd{N^2}$ and it is therefore comparable to that of computational regularization methods implementable on classical machines (which, in addition, provide the full solution vector).\\

Finally, it is worth comparing the QLSA to classical parallel methods for matrix inversion. In the parallel model of computation~\cite{heller1978survey}, inverting an $N\times N$ matrix takes $\Ord{\text{log}^2(N)}$ computational steps using a number of processors which is of order $\text{poly}(N)$ (a crude upper bound of $\Ord{N^4}$ is given by~\cite{csanky1976fast}). Although the parallel model of computation does not resemble the actual behaviour of parallel machines, it can be a fair comparison considering that quantum computers might also face connectivity issues and hence communication overheads among the qubits. In  particular when exponentially large amounts of error-corrected qubits are required, as with the QRAM, it is likely that latency issues arise.\\

To conclude, the QLSA is a logarithmic time quantum algorithm for matrix inversion, a task arising in many learning problems. However, a number of caveats that include the requirement of a logarithmic access time to the memory and the impossibility of retrieving the solution vector with one measurement, lead to question whether classical or parallel algorithms that make use of the same assumptions obtain similar, or better, runtimes. In this respect, experimental implementations will greatly contribute to asses the true potential of these methods in realistic scenarios.

\subsection*{Quantum singular values estimation}

The \textit{singular value decomposition} (SVD) of a $M \times N$, rank $r$ matrix $A$ is a factorisation of the form 
$A = U \Sigma V^{\dagger}$, where $U$ and $V$ are, respectively, $M \times M$ and $N \times N$ unitary matrices and $\Sigma$ is a $M \times N$ diagonal matrix 
with $r$ positive entries $\sigma_1 , \ldots, \sigma_r$ which are called the \textit{singular values} of $A$.\\ 

Singular value estimation is a fundamental tool in many computational problems and applications ranging from matrix inversion for linear regression to matrix approximation~\cite{golub2012matrix}. It is also of particular interest for problems of dimensionality reduction like \textit{principal component analysis} (PCA)~\cite{jolliffe1986principal}. Classically, finding such a decomposition is computationally expensive, and for $M > N$ it takes $\Ord{M N^2}$~\cite{trefethen1997numerical}.\\ 

Prakash and Kerenidis~\cite{prakash2014quantum, kerenidis2017quantum} introduced the \textit{quantum singular value estimation} (QSVE) algorithm, based on Szegedy's work on quantum walks~\cite{szegedy2004quantum}, which runs in time $\Ord{ ||A||_F \log{MN}/\epsilon }$. Their algorithm returns an estimate of the singular values $\tilde{\sigma_i}$ such that $| \sigma_i - \tilde{\sigma}_i| \leq \epsilon$. As for the QLSA, the QSVE algorithm outputs the singular values  in quantum superposition. As such, in order to read out all the $r$-values, the algorithm must be run $\Ord{N\log N}$, times thus destroying any exponential speedup. However, it is still possible to construct useful applications of the QSVE algorithm. For example, \cite{kerenidis2017quantum} proposed a recommendation system which runs in
$\Ord{\text{poly}(r)\text{poly}\log(MN)}$ (assuming a good $r$-rank approximation of the preference matrix). \\

We note that the QSVE algorithm requires an oracle that can prepare quantum states that encode the rows and the columns of the matrix $A$ in polylogarithmic time. It is possible to implement this oracle with the QRAM, and hence it will inherit the caveats discussed in Section~\ref{sec:data}.\\

An alternative method for quantum singular value estimation, has been proposed by Lloyd, Mohseni, and Rebentrost~\cite{lloyd2014pca}. The scaling of this algorithm is quadratically worse in $\epsilon$ but the requirements on the memory structure are less stringent than in~\cite{kerenidis2017quantum}. This is advantageous in some applications, like analysing the principal components of kernel matrices~\cite{rebentrost2014quantum}.\\ 

\section{Quantum methods for sampling}
\label{sec:sampling}

Many learning problems of practical interest, as, for example, exact inference in graphical models, are intractable with exact methods. We discuss in detail hard learning problems in Section~\ref{sec:hard}. Sampling methods are a common technique to compute approximations to these intractable quantities \cite{neal1993probabilistic}. There is a rich literature on sampling methods \cite{neal2001annealed,gilks1992adaptive,propp1996exact,doucet2001introduction,del2006sequential}. The most commonly used ones are Monte Carlo methods and in particular the MCMC. The quantum algorithms discussed in this section are devoted to speed up MCMC methods.\\

MCMC methods~\cite{sinclair1993markov}, like Gibbs Sampling or the Metropolis algorithm, allow to sample from a probability distribution $\Pi$ defined over a state space using a Markov chain that after a number of steps converges to the desired distribution (in practice one will only reach a distribution which is $\epsilon$-close). The number of steps $\tau$ required to converge to $\Pi$ is referred to as the \textit{mixing time}. Estimating the mixing time can be reduced to bounding the spectral gap $\delta$, which is the distance between the largest and the second largest eigenvalue of a stochastic map that evolves the Markov chain. The mixing time satisfies the inequality $\tau \geq \frac{1}{2 \delta} \log{(2 \epsilon)}^{-1}$ and it is possible to show~\cite{sinclair1993markov,wocjan2009quantum} that for the classical MCMC algorithm, $\tau$ is of the order $\Ord{1/(\delta \log{(1/\Pi^*)})}$, where $\Pi^*$ is the minimum value of $\Pi$.\\

Recently, there has been a significant interest in quantum algorithms that allow to speedup the simulations of the stochastic processes used in MCMC. A common feature of these algorithms is a quadratic speedup in terms of spectral gap, inverse temperature, desired precision or the hitting time. Advances in this field include algorithms for thermal Gibbs state preparation~\cite{somma2008quantum,poulin2009sampling,chiang2010quantum,bilgin2010preparing,temme2011quantum,schwarz2012preparing,chowdhury2016quantum} which provide polynomial speedups in various parameters, such as the spectral gap. Other methods have introduced the concept of quantum hitting time of a quantum walk~\cite{szegedy2004quantum,ambainis2005coins,krovi2006hitting,krovi2010adiabatic,magniez2011search,chowdhury2016quantum}. In this framework it is possible to obtain an polynomial speedup with respect to most classical variants (this can be exponential for the hitting time).
A number of other algorithms accelerate classical Monte Carlo methods applied to the estimation of quantities such as expectation values and partition functions, which play a major role in physics~\cite{montanaro2016quantum,knill2007optimal,chowdhury2016quantum}.\\


\section{Quantum optimisation}
\label{sec:opt}

As discussed in Section~\ref{sec:setting}, optimisation methods are a fundamental building block of many machine learning algorithms. Quantum computation provides tools to solve two broad classes of optimisation problem: semidefinite programming and constraint satisfaction problems.

\subsection*{Quantum algorithms for semidefinite programming}

Semidefinite programming~\cite{vandenberghe1996semidefinite, grotschel2012geometric} is a framework  for solving certain types of convex optimisation problems. Semidefinite programs find widespread applications in ML \cite{lanckriet2004learning,weinberger2007graph,jacob2009group}. In a semidefinite program (SDP) the objective is to minimise a linear function of a $N \times N$ positive semidefinite matrix $X$ over an affine space defined by a set of $m$ constraints. The best known classical SDP-solvers~\cite{lee2015faster} runs in time $\mathcal{O}(m(m^2+n^\omega + mns)\mathrm{log}^{O(1)}(mnR/\epsilon))$, where $\epsilon$ is an approximation parameter, $\omega \in [2,2.373)$ is the optimal exponent for matrix multiplication, $s$ is the sparsity of $A$, and $R$ is an upper bound on the trace of an optimal~$X$.\\

Based on a classical algorithm to solve SDPs by Arora and Kale~\cite{arora2007combinatorial}, that has a runtime of $\tilde{\mathcal O} (nms\left(Rr / \epsilon \right)^4+ns\left(Rr/ \epsilon \right)^7)$, where $r$ is an upper bound on the sum of the entries of the optimal solution to the dual problem, in 2016 Brand\~{a}o and Svore~\cite{brandao2016quantum} developed a quantum algorithm for semidefinite programs that is quadratically faster in $m$ and $n$. The dependence on the error parameters of this result has been improved in~\cite{van2017quantum}. In this work the authors obtain a final scaling of $\widetilde{\mathcal O}(\sqrt{mn}s^2 (Rr/\epsilon)^{8})$.\\

The main problem of these quantum algorithms is that the dependence on $R,r,s$ and $1/\epsilon$ is considerably worse than in~\cite{arora2007combinatorial}. This quantum algorithm thus provides a speed-up only in situations where $R,r,s,1/\epsilon$ are fairly small compared to $mn$ and, to date, it is unclear if there are interesting examples of SDPs with these features (for more details see~\cite{van2017quantum}).

\subsection*{Quantum algorithms for constraint satisfaction problems}

In a \textit{constraint satisfaction problem} (CSP), we are given a set of variables, a collection of constraints, and a list of possible assignments to each variable~\cite{creignou2001complexity}. The task is to find values of the variables that satisfy every constraint. This setting prompts to exact and approximate cases. For many families of CSPs efficient algorithms are unlikely to exist. Two quantum algorithms are known for CSPs: the quantum approximate optimisation algorithm and the quantum adiabatic algorithm. Due to its generality and a profoundly different way of exploiting quantum evolution, the latter algorithm is also regarded as an independent computational model called \textit{adiabatic quantum computation} (AQC). We will provide a brief introduction to AQC in the following paragraphs.
 
\subsubsection*{The quantum approximate optimisation algorithm}

The \textit{quantum approximate optimisation algorithm} (QAOA) developed in 2014 by Farhi, Goldstone and Gutman is a quantum method to approximate CSPs~\cite{farhi2014quantum}. The algorithm depends on an integer parameter $p \geq 1$ and the approximation improves as $p$ increases. For small values of $p$ the QAOA algorithm can be implemented on a shallow circuit. As argued~\cite{farhi2017quantum} this feature makes the QAOA algorithm a good candidate for first generation quantum hardware.\\

For certain combinatorial optimisation problems the QAOA algorithm can give approximation ratios that are better than what can be achieved by random sampling~\cite{farhi2014quantum} but worse than the best classical solvers. In specific instances of MAX-\textit{k}XOR the QAOA algorithm with $p = 1$ was believed to outperform the best classical solver~\cite{farhi2014applied}. This sparked further research in the classical community and Barak \textit{et al.} designed a classical algorithm able to outperform the quantum scaling~\cite{barak2015beating}.

\subsubsection*{The quantum adiabatic algorithm}

The \textit{quantum adiabatic algorithm} (QAA)~\cite{farhi2000quantum} is an optimisation method that operates in the adiabatic model of quantum computation. The QAA can be thought of as a quantum analogue of simulated annealing~\cite{kirkpatrick1984optimization}. The algorithm encodes the solution to a computational problem in the unknown ground state of a quantum system (usually an Ising spin glass Hamiltonian). By starting off in the ground state of a known and easy to implement Hamiltonian, the QAA exploits a slow, time-dependent, Hamiltonian dynamics to obtain the solution to the problem. If the evolution is slow enough, the quantum adiabatic theorem~\cite{messiah1958quantum} guarantees that the system will reach the desired ground state. If the energy barriers have specific configurations (e.g. tall and narrow) and the energy gap between the ground state and the first excited state remains large enough, the algorithm can obtain significant speedups over classical simulated annealing~\cite{reichardt2004quantum, crosson2016simulated}. \\

Although QAA and AQC are usually considered synonyms in the literature we shall keep the two concepts distinct as to mark the difference between the computational model and the algorithm. Another name which is frequently used in the literature as synonym of QAA and AQC is \textit{quantum annealing} (QA). Although there is not a clear consensus in the literature over the differences between these three concepts, we refer to QA only when the adiabatic evolution occurs at non-zero temperature.\\

Aharonov \textit{et al.}~\cite{aharonov2008adiabatic} showed that AQC is universal for quantum computation, \textit{i.e.} it is capable of solving any computational problem that can be solved by a quantum computer. Although it is clearly possible to encode \NP-hard problems~\cite{barahona1982computational}, quantum mechanics is not expected to solve these in polynomial time (however the scaling constants of the quantum algorithm might be smaller). Finally, it is important to note that the adiabatic algorithm lacks worst case upper bounds on its runtime. Its performance has been analysed with numerical experiments \cite{farhi2000numerical,farhi2001quantum,farhi2012performance,hen2011exponential,young2008size,young2010first}. However, these are limited to small size systems and only running the algorithm on actual hardware will be able to determine the strength of this approach. 

\section{Quantum neural networks}
\label{sec:QNN}

The term \textit{artificial neural network} (ANN) denotes a variety of models which have been widely applied in classification, regression, compression, generative modelling and statistical inference. Their unifying characteristic is the alternation of linear operations with, usually preselected, non-linear transformations (e.g. sigmoid functions) in a potentially hierarchical fashion. \\

While in the last decade neural networks have proved successful in many applications, fundamental questions concerning their success remain largely unanswered: are there any formal guarantees concerning their optimisation and the predictions they return? How do they achieve good generalisation performance despite the capacity to completely overfit the training data?\\

Artificial neural networks have been extensively studied in the QML literature. The major research trends have focused on accelerating the training of classical models and on the development of networks where all the constituent elements, from the single neurons to the training algorithms, are executed on a quantum computer (a so called \textit{quantum neural network}). The first works on quantum neural networks appeared in the 90's~\cite{kak1995quantum} and a number of papers have been published on the topic. However, it is worth noticing that the field has not reached a level of scientific maturity comparable to the other areas of QML discussed in this review. Possible reasons for the difficulties encountered in making progress in this area can be traced to the inherent differences between the linearity of quantum mechanics and the critical role played by non-linear elements in ANNs or the fast developments occurring in the field of classical ANNs.\\

The literature on accelerated training of NNs using quantum resources has mainly focused on \textit{restricted Boltzmann machines} (RBMs). RBMs~\cite{smolensky1986information} are generative models (\textit{i.e.} models that allow to generate new observational data based on prior observations) that are particularly apt to be studied from a quantum perspective due to their strong connections with the Ising model. It has been shown that computing the log-likelihood and sampling from an RBM is computationally hard~\cite{long2010restricted}. \textit{Markov Chain Monte Carlo} (MCMC) methods are the standard techniques used to overcome these difficulties. Nonetheless, even with MCMC the cost of drawing samples can be high~\cite{dumoulin2014challenges} for models with several neurons. Quantum resources can help to reduce the training cost. \\

There are two main classes of quantum techniques to train RBMs. The first one is based on methods from quantum linear algebra (discussed in Section~\ref{sec:linear}) and quantum sampling (discussed in Section~\ref{sec:sampling}). Wiebe, Kapoor and Svore~\cite{wiebe2014quantum} developed two algorithms to efficiently train a RBM based on amplitude amplification~\cite{brassard2002quantum} and quantum Gibbs sampling. These obtain a quadratic improvement in the number of examples required to train the RBM but, the scaling of the algorithm is quadratically worse in the number of edges than contrastive divergence~\cite{hinton2002training}. A further advantage of the approach proposed in~\cite{wiebe2014quantum} is that it can be used to train full Boltzmann machines (a classical version of this algorithm has also been proposed~\cite{wiebe2015quantum}). A full BM is a type of Boltzmann machine where the neurons correspond to the nodes of a complete graph (\textit{i.e.} they are fully connected). Although full BMs have an higher number of parameters with respect to RBMs, they are not used in practice due to the high computational cost of training and, to date, the true potential of large scale, full BMs is not known.\\

The second direction to training RBMs is based on quantum annealing, a model of quantum computation that encodes problems in the energy function of an Ising model (quantum annealing will be discussed in Section~\ref{sec:opt}). Specifically,~\cite{adachi2015application, denil2011toward} make use of the spin configurations generated by a quantum annealer to draw Gibbs samples that can be then used to train a RBM. These types of physical implementations of RBMs present several challenges. Benedetti and co-authors~\cite{benedetti2016estimation} pointed out the difficulties in determining the effective temperature of the physical machine. In order to overcome this problem they introduced an algorithm to estimate the effective temperature and benchmarked the performance of a physical device on some simple problems. A second critical analysis of quantum training of RBMs was conducted by Dumoulin \textit{et al.}~\cite{dumoulin2014challenges}. Here the authors showed with numerical models how the limitations that the first generations quantum machines are likely to have in terms of noise, connectivity and parameters tuning, severely limit the applicability of quantum methods.\\

An hybrid approach between training ANNs and a fully quantum neural network is the quantum Boltzmann machine proposed by Amin \textit{et al.}~\cite{amin2016quantum}. In this model the standard RBM energy function gains a purely quantum term (\textit{i.e.} off diagonal) that, according to the authors, allows to model a richer class of problems (\textit{i.e.} problems that would otherwise be hard to model classically like quantum states). Whether these models can provide any advantage for classical tasks is unknown. Kieferova and Wiebe~\cite{kieferova2016tomography} suggest quantum Boltzmann machines could provide advantages for tasks like reconstructing the density matrix of a quantum state from a set of measurements (this operation is known in the quantum information literature as \textit{quantum state tomography}).\\

Although there is no consensus on the defining features of a quantum artificial neural network, the last two decades have seen a variety of works that attempted to build networks whose elements and updating rules are based solely on the laws of quantum mechanics. The review by Schuld, Sinayisky and Petruccione~\cite{schuld2014quest} provides a critical overview of the different strategies employed to build a quantum ANN and highlights how most of the approaches do not meet the requirements of what can be reasonably defined as a quantum ANN. In particular, most of the papers surveyed by Schuld \textit{et al.} failed to reproduce basic features of ANNs (for example, the the attractor dynamics in Hopfield networks). On the other side, it can be argued that the single greatest challenge to a quantum ANN is that the quantum mechanics is linear but ANNs require non linearities~\cite{cybenko1989approximation}.\\

Recently, two similar proposals~\cite{wan2016quantum, romero2016quantum} have overcome the problem of modelling non-linearities by using measurements and introducing a number of overhead qubits in the input and output of each node of the network. However these models still lack some important features of a fully quantum ANN. For example, the models parameters remain classical, and it is not possible to prove that the models can converge with a polynomial number of iterations. The authors of the papers acknowledge that, in their present forms, the most likely applications of these models appear to be learning quantum objects rather than enhancing the learning of classical data. Finally, we note that, to date, there are no attempts to model non linearities directly on the amplitudes.\\

\section{Learning with noise}
\label{sec:noise}

Noise can play different, potentially beneficial, roles in learning problems.
In a classical setting, it has been shown that noise can alleviate two of the most common model-fitting issues: local optima and generalisation performance. Perturbing gradients can help with the former by ``jumping out'' of local optima, whereas perturbing training inputs or outputs can improve the latter.\\

The possibility of exploiting advantageously the effects of noise is particularly interesting in the context of quantum computation. Early quantum computers are expected to have too few qubits to implement full error correction and the community is actively looking for problems where noise not only does not destroy the computation but can play a beneficial role. \\

The analysis of noisy learning problem from a quantum perspective becomes particularly promising in selected cases. As we will discuss in this section, quantum resources allow to solve efficiently noisy learning problems that would be otherwise classically hard. Although few results are known in this area, further research in this direction might provide new cases of a separation between the classical and quantum case in a learning setting.\\

The goal of this section is to inspire future research aimed at understanding how quantum learners behave in noisy settings. We begin by reviewing for the quantum scientists a number of classical problems in machine learning that benefit from noise. We proceed with a brief introduction to standard ways of modelling errors in quantum computing aimed at machine learning practitioners. We conclude by discussing problems where quantum resources allow to perform tasks that would be otherwise hard for a classical learner. 

\subsection*{Classical learning can benefit from noise}

\subsubsection*{Noisy inputs}

The first direct link between the addition of noise to the training inputs \((x_i)_{i=1}^n\) and Tikhonov regularisation was drawn in~\cite{bishop1995training}. Here, it is shown that optimising a feed-forward neural network to minimise the squared error on noisy inputs is equivalent (up to the order of the noise variance) to minimising the squared error with Tikhonov regularisation on noiseless inputs. \\

Intuitively, this form of regularisation forces the gradient of the neural network output \(f(x)\) with respect to the input \(x\) to be small, essentially constraining the learned function to vary slowly with \(x\): neighbouring inputs are encouraged to have similar outputs. \\

An~\cite{an1996effects} also investigated the effects of adding noise to inputs, outputs, weights and weight updates in neural networks and observed that input (and sometimes weight) noise can, in some settings, improve generalisation performance. 

\subsubsection*{Noisy parameter updates}

More recently, in~\cite{neelakantan2015adding}, the addition of annealed i.i.d Gaussian noise to the gradients has been empirically shown to help in optimising complex neural network models.
Indeed, stochasticity in the optimisation process can also derive from evaluating gradients of the objective function with respect to randomly selected subsets of the training points (as in \emph{stochastic gradient descent}). This can be intuitively compared to simulated annealing~\cite{bottou1991stochastic} since the natural variability in these ``partial'' gradients can help escape local optima (and saddle points) and the (decreasing) gradient step size can be directly compared to the annealing temperature.\\

The addition of noise to the update of model parameters was also adopted in \cite{welling2011bayesian}. There, as well as using random subsets of training points to evaluate gradients, at each iteration the parameter update is perturbed with Gaussian noise (with variance equal to the decreasing step size). After the initial stochastic optimisation phase, it can be shown that this method, under specific conditions, will start generating samples from the posterior distribution over model parameters, allowing us to quantify model uncertainty and avoid overfitting at no extra computational cost.\\

\subsubsection*{Noisy outputs}

In Gaussian process regression \cite{rasmussen2006}, on the other hand, noise in the training outputs \((y_i)_{i=1}^n\) helps avoid the inversion of an otherwise potentially ill-conditioned kernel covariance matrix $K$. Assuming additive isotropic Gaussian noise (with variance \(\sigma^2\)), to evaluate model predictions, we only ever need to invert a matrix of the form \(K + \sigma^2 I\). This can be practical as the kernel matrix is singular or ill-conditioned whenever training inputs are repeated or are very close in the Hilbert space associated with the kernel covariance function.\\

Finally, when training \textit{generative adversarial networks} (GANs, \cite{goodfellow2014generative}) it has been shown that an overconfident ``discriminator'' can hinder learning in the ``generator''. In GANs in fact, a generative model (the ``generator'') is trained by attempting to ``deceive'' a ``discriminator'' model into classifying the generated images as coming from the true data distribution. However, especially early on in training, there might be little overlap in the support of the data distribution and the generator. This can result in the discriminator predicting labels with very high confidence and, as well as potentially overfitting, in making the discrimination decision depend very weakly on the generator's parameters. 
To address this issue, labels (\textit{i.e.} true, fake) can be ``fuzzied''. Specifically, for each training point, the discriminator will assume that all \(K\) labels have probability at least $\epsilon/K$ of occurring, with the true label having probability \((1 - \epsilon) + \epsilon/K\). This corresponds to assuming that with probability \(\frac{\epsilon}{K}\) labels are sampled at random and, indeed, labels can just be flipped randomly in practice. Effectively, this keeps the model from becoming too confident in its predictions by making it suboptimal to shift all the probability mass on the true label. This technique is called label-smoothing~\cite{szegedy2016rethinking} and it has been shown to help retain training signal for the generator \cite{salimans2016improved}, as well as increasing robustness of classifiers to adversarial examples \cite{warde2016perturbation}.\\

\subsection*{A classical/ quantum separation in learning under noise}

In order to address learning under noise in a quantum setting it is necessary to discuss what type of noise affects quantum computers. The works by Preskill~\cite{preskill1998fault} and Breuer and Petruccione~\cite{breuer2002theory} cover the topic extensively. A simple model of quantum errors, usually employed in numerical simulation of noisy quantum devices, makes use of a weighted combination of two kinds of error: bit flips and phase flips. We can justify this simple type of modelling because, in the most common error correcting codes, errors are detected by projecting more complex errors into convex combinations of bit and phase flips. Given a quantum state $\psi = \alpha_0 e_0 + \alpha_1 e_1$, a bit flip error turns the state into $\tilde{\psi} = \alpha_0 e_1 + \alpha_1 e_0$. Similarly, a phase flip error changes the relative phase of a quantum state, \textit{i.e.} the resulting state is  $\tilde{\psi} = \alpha_0 e_0 - \alpha_1 e_1$. More complex and realistic models of errors include amplitude damping, leakage to higher levels, and loss .\\

Many authors have studied how noise affects the learnability of a function in the quantum setting. The already mentioned work by Bshouty and Jackson~\cite{bshouty1998learning} showed that DNF formulas can be efficiently learned under the uniform distribution using a quantum example oracle. This contrasts with the classical case (although proved in the statistical query model of learning) where Blum and co-authors showed that DNF are not learnable under noise with respect to the uniform distribution~\cite{blum1994weakly}. \\

Another result that points to a separation between classical and quantum for a noisy learning problem has been recently proved by Cross, Smith, and Smolin~\cite{cross2015quantum}. In this case, it is discussed the learnability of parity functions under noise. It is widely believed that learning parity function under noise is not classically efficient~\cite{lyubashevsky2005parity} and the best classical algorithm run in subexponential, but superpolynomial, time. Furthermore, the problem is an average case version of the \NP-hard problem of decoding a linear code~\cite{berlekamp1978inherent}, which is also known to be hard to approximate~\cite{haastad2001some}. Both the classical and quantum problem are easy without noise. In~\cite{cross2015quantum} was shown that in the quantum PAC model parity functions can be learned efficiently under the uniform distribution (with logarithmic overhead over the noiseless runtime). Their results have been generalised to linear functions and to more complex error models by Grilo and Kerenidis~\cite{grilo2017learning}. \\

To summarise, in this section we surveyed a number of classical results showing that noise in the inputs, outputs or in the parameters can have positive effects on  learning algorithms. It would be interesting to investigate whether the type of noise encountered in quantum systems has a similar distribution and structure to the one commonly encountered in classical settings. In this case machine learning algorithms would become ideally suited to run on non fault-tolerant quantum hardware. Finally, further research is needed to identify new, noisy, problems that only a learner equipped with quantum resources can solve.

\section{Computationally hard problems in machine learning}
\label{sec:hard}

Algorithms whose runtime is upper bounded by a polynomial function of $N$ are said to be \textit{efficient}. Problems for which there exists an efficient algorithm are \textit{easy}. Conversely, \textit{hard} problems are those where no polynomial algorithm is known. An important class of easy problem is called \P. The class of problems that are efficiently solvable by a quantum computer includes some problems that are not known to be in \P.\\


The quantum algorithms surveyed in this review speed up efficient classical algorithms. Two types of speedups are obtained: polynomial or exponential. Polynomial speedups, although important from a practical point of view, do not prove that quantum computers are able to turn hard learning problems into easy ones. On the other hand, exponential speedups of algorithms that are already efficient face important challenges. Indeed, as we have seen for the matrix inversion algorithm discussed in Section~\ref{sec:linear}, quantum algorithms for the analysis of classical data running in logarithmic time require an equally fast access to the memory. This can be obtained using a QRAM that, however, presents a number of issues (see Section~\ref{sec:data}).\\

In order to achieve an exponential speedup despite the computational costs arising from accessing the memory we are restricted to hard algorithms. This is because, for these algorithms, the polynomial time construction of the quantum state that encodes the dataset does not dominate over the speedup. We discussed an example with such a property: the learnability of DNF formulas (Section~\ref{sec:better}). Classically, the best algorithm for learning DNFs runs in superpolynomial time. With quantum resources we can learn the same problem polynomially. Although these types of learning problems have limited practical applications, they suggest that an exponential separation between classical and quantum models of learning might hold in real world problems. \\

In this section, we present a number of problems in machine learning that are believed to be computationally hard and are receiving considerable interest in the classical community. We do not expect that these problems, some of which are \NP-hard, can be solved efficiently with a quantum computer. Recall that \NP-hard is a class of problems for which there is strong evidence of a separation with \P~\cite{fortnow2009status}. Our hope is to spark interest in the search for hard problems in machine learning with the kind of structure (see Section~\ref{sec:essential}) that can be exploited by quantum computers. We also decided to include problems that are not hard in the computational complexity sense but whose high degree polynomials runtime make them intractable. For these cases, where slow (\textit{i.e.} polynomial) memory access times can still be tolerable, even polynomial speedups might be of great practical relevance.

\subsection*{Tensor factorisation} 
As modern datasets grow not only in terms sheer dimension but also in the complexity of the structures required to store such data (e.g. multi-modal data, social networks, recommender systems, and relational data~\cite{suchanek2007yago,dong2014knowledge,carlson2010toward}), it becomes ever more critical to device methods able to distil concise and interpretable representations of this information. Tensor models offer a powerful way to address these learning problems. For instance, tensors naturally generalize the concept of adjacency matrix for multi relational graphs \cite{getoor2007introduction}. However, given the intrinsic multi dimensional nature of these objects, tensor based learning problems are typically computationally hard and require large amounts of memory, therefore become quickly impractical in most applications. To this end, finding low rank approximations of tensors (or more generally multi-linear operators), a natural generalisation of the problem of matrix factorisation (see \cite{kolda2009tensor} and references therein), has recently received significant attention from the fields of machine learning, inverse problems and compressive sensing. However, while for the matrix case the problem is amenable to efficient computations, moving to higher orders becomes significantly challenging. Indeed, in contrast to its matrix counterpart, low rank tensor factorisation, even when relaxed to a nuclear norm regularised optimization problem, has been recently shown to be \NP-hard \cite{hillar2013most}. Approaches have attempted to circumvent these issues by considering further relaxation of the factorisation problem \cite{mu2014square,richard2014statistical,romera2013new,romera2013multilinear,signoretto2014learning,gandy2011tensor}, but to this day a standard solution has yet to be proposed.

\subsection*{Submodular problems} 

Recently, several machine learning problems have been addressed via submodular optimization. Examples of such applications are very diverse, such as document
summarisation \cite{lin2011class}, social networks \cite{kempe2003maximizing}, or clustering \cite{narasimhan2007local} to name a few. Submodularity characterises a family of discrete optimisation problems, typically entailing cost functions on sets, in which the target functional exhibits a structure akin to that of convexity (or rather concavity) for continuous functions. We refer to \cite{bach2015submodular} for an in-depth introduction on the topic. For many submodular problems it is possible to identify a corresponding convex problem via the so-called {\em Lov\`asz} extension~\cite{lovasz1982submodular}. As a consequence, such problems can be solved using convex optimisation methods, leading to efficient learning algorithms. However, for a wide range of these problems, the corresponding computational complexity, albeit polynomial, is of high order (e.g. $\mathcal O(n^5)$ with respect to the number $n$ of the parameters, see for instance \cite{schrijver2000combinatorial,iwata2001combinatorial,orlin2009faster}), making them remarkably slow in practice. In this sense, an exponential (or even polynomial) decrease in the number of computations to solve a submodular problem, analogous to the one observed for fast linear algebra using quantum algorithms, could be key to tackle practical applications.





\subsection*{Inference in graphical models}


Probabilistic models in machine learning can be encoded in graphs. Graphical models of particular use are Bayesian networks~\cite{pearl1985bayesian} and Markov random fields~\cite{kindermann1980markov}: directed acyclic and undirected graphs, respectively, where nodes represent random variables and edges denote dependence between variables. Operations like marginalisation and conditioning can be performed by algorithms taking into account the specific connectivity of the given graph (\textit{i.e.} message passing). While this offers a general framework for inference (\textit{i.e.} evaluating the distribution of latent variables conditioned on observed ones), it has been shown, by reduction to Boolean satisfiability \cite{cooper1990computational}, that exact inference in these networks is \NP-hard and that evaluating the normalising constant \(Z\) (or partition function) for the joint distribution is in \#\P~(a family of hard counting problems).\\


\section{Conclusions and outlook}
\label{sec:outlook}

In this review, we surveyed a number of different quantum methods to tackle learning problems. 
Despite a number of promising results, the theoretical evidence presented in the current literature does not yet allow us to conclude that quantum techniques can obtain an exponential advantage in a realistic learning setting. Even in the case of quantum algorithms for linear algebra, where rigorous guarantees are already available, issues related to data access and restrictions on the types of problems that can be solved might hinder their performance in practice.
In fact, near future advances in quantum hardware development will be important to empirically assess the true potential of these techniques. In this regard, we note how the great majority of the quantum machine learning literature has been developed within the quantum community. We believe that further advances in the field will only come after significant interactions between the two communities. For this reason, we tried to structure the review in a way that presents the different topics in a way familiar to both quantum scientists and machine learning researchers. In order to achieve this goal we put great emphasis on the computational aspects of machine learning. Although this perspective has the obvious advantage of providing an agile way for discussing quantum algorithms (that mostly focus on accelerating the runtime with respect to their classical counterparts), the reader should keep in mind that statistical problems (like determining the generalisation performance of an algorithm) are equally relevant. The approach taken in this review has also left some interesting papers aside (see for example~\cite{lloyd2014quantum}). We invite the reader to consult~\cite{schuld2014introduction,adcock2015advances, biamonte2016quantum} for a review that includes these works.\\

In Section~\ref{sec:setting} we discussed how the computational cost represents one of the major challenges for the future of machine learning. In particular, polynomial scaling in the number of data points might not be adequate in the age of large scale machine learning. The quantum algorithms presented here allow to reduce the complexity of some, currently used, regularisation methods. We classified the quantum approaches into four main categories: linear algebra, neural networks, sampling, and optimisation. The quantum machine learning algorithms based on linear algebra subroutines are those that promise the greatest computational advantages (\textit{i.e.} exponential). However, it is not clear whether fundamental limitations related to how quickly these algorithms need to access the memory might compromise their ability to speed up the analysis of classical data. Quantum methods for training neural networks, for sampling, and for optimisation, provide so far mostly quadratic advantages and some of these might be implementable on first generation quantum computers. Unfortunately, the theoretical framework on which they are based is not yet well established (e.g. the quantum Boltzmann machines described in Section~\ref{sec:QNN}) and only practical experiments will determine their true performance.\\

To summarise, the works surveyed in this review, including the theoretical evidence presented in Section~\ref{sec:better}, suggest the possibility of a quantum speedup for some machine learning problems. However, the extent of these speedups, and consequently the impact of these methods on practical problems, remains an open question.\\

We identified a number of promising directions for the field. First, exploring the tradeoffs between noise, generalisation performance and hardness in a quantum context (Section~\ref{sec:noise}). This is particularly interesting for first generation quantum hardware that most likely will not be fault-tolerant. 
Second, deepening our understanding of how quantum resources can affect sample and time complexity, even for problems that are already known to be efficient. 
Significant work has already been done but some areas like statistical learning theory are yet to receive a thorough analysis in a quantum context. Third, determining whether a QRAM of the size required to handle large datasets can be constructed on a physical device (Section~\ref{sec:data}). Fourth, understanding whether there exist non polynomial problems in machine learning that can be tackled efficiently using quantum resources (Section~\ref{sec:hard}). This direction is arguably the most relevant for finding quantum algorithms capable of demonstrating an uncontroversial speedup in a learning context, and this is indeed the general quest of quantum computation.\\

\subsection*{Data accessibility}

This paper has no data.

\subsection*{Competing interests}

The authors declare no competing interests.

\subsection*{Authors' contributions}

AR and LW conceived the project. All authors contributed to the literature review. All authors wrote the manuscript. 

\subsection*{Acknowledgements}

We thank Scott Aaronson, David Barber, Marcello Benedetti, Fernando Brand\~{a}o, Dan Brown, Carlos Gonz\'alez-Guill\'en, Joshua Lockhart, and Alessandro Rudi for helpful comments on the manuscript. 

\subsection*{Funding statement}

AI is supported by the Cambridge-Tuebingen Fellowship and the Qualcomm Innovation Fellowship. AR is supported by an EPSRC DTP Scholarship and by QinetiQ. CC and MP are supported by EPSRC. SS is supported by The Royal Society, EPSRC, Innovate UK, Cambridge Quantum Computing, and the National Natural Science Foundation of China. 

\printbibliography

\end{document}